\documentclass[preprint,12pt,authoryear]{elsarticle}
\usepackage{natbib}

\usepackage{amssymb}
\usepackage{amsmath}

\usepackage[pdftex]{hyperref}
\usepackage{float}
\usepackage{subcaption}
\usepackage{booktabs}
\usepackage{longtable}

\journal{Computers, Environment and Urban Systems}

\begin{document}

\begin{frontmatter}



\title{\texttt{rt2gtfs}: A scalable framework for correcting public transport timetables using real-time data for accessibility analysis} 


\author{Zihao Chen} 

\affiliation{organization={University of Exeter},
            addressline={North Park Road}, 
            city={Exeter},
            state={Devon},
            postcode={EX4 4QF}, 
            country={United Kingdom}}

\author{Federico Botta} 

\affiliation{organization={University of Exeter},
            addressline={North Park Road}, 
            city={Exeter},
            state={Devon},
            postcode={EX4 4QF}, 
            country={United Kingdom}}

\begin{abstract}
Travel time is a fundamental component of accessibility measurement, yet most accessibility analyses rely on static timetable data that assume public transport services operate exactly as scheduled. Such representations overlook the substantial variability in travel times arising from operational conditions and service disruptions. In this paper, we present \texttt{rt2gtfs}, an open-source Python package for reconstructing empirical public transport timetables from high-frequency vehicle location data. The package provides a configurable and scalable workflow for collecting GTFS-Realtime vehicle position feeds from the UK Bus Open Data Service (BODS), matching observed vehicle locations to scheduled GTFS trips and stops, inferring stop-level arrival and departure times, and exporting corrected GTFS format timetable bundles. Using national-scale real-time bus data feeds from BODS, we demonstrate how \texttt{rt2gtfs} can be used to generate observed timetables suitable for routing and origin-destination travel time calculation, as well as accessibility analysis. By packaging the framework as reusable software, this work supports more reproducible and realistic accessibility analysis and provides a practical tool for researchers and practitioners seeking to incorporate observed public transport performance into transport planning.
\end{abstract}




\end{frontmatter}



\section{Introduction}

Fair access to essential services such as healthcare, education and grocery shops is fundamental to spatial justice and transport equity. Public transport plays a particularly important role in enabling such access, especially for populations without access to private vehicles. Poor public transport access to services can lead to social exclusion, reduced wellbeing, and diminished life opportunities \citep{farber_time-use_2011, lucas_transport_2012, stanley_place-based_2022}, resulting in what are often termed transport deserts, which are places where opportunities exist but are effectively out of reach \citep{jiao_transit_2013}. Consequently, measuring public transport accessibility has become a central task in transport geography and planning research, informing both academic understanding of spatial inequalities and policy interventions aimed at improving service provision. 

Substantial research has sought to measure accessibility to services, with some having been conducted at national or global scales \citep{weiss_global_2018, weiss_global_2020,verduzco_torres_public_2024, ye_national-scale_2024}. However, they typically focus on a single time point or rely on average travel times derived from static published timetables, overlooking temporal variability and operational uncertainty. This approach implicitly overlooks the fact that the same journey could take different amount of time to complete at different times of day or day of week (known as Travel Time Variability, TTV), and assumes that services run as planned without delay or cancellation (known as Travel Time Inaccuracy, TTI), which often leads to miscalculation of accessibility \citep{liu_realizable_2023, lee_social_2023, chen_measuring_2017, javanmard_using_2025}. In reality, public transport operations, particularly road-based ones, are affected by congestion, breakdowns, weather conditions, and many other disruptions that vary systematically by place and time of day. Ignoring these dynamics can obscure the temporal dimension of accessibility and potentially misrepresent the geography of transport-related inequalities. Even without considering real-world operational disruptions, our previous work \citep{chen_geography_2025} identified variability in travel times across the day when journeys were modelled with different departure times using timetable data alone; this variability also differed temporally and geographically. Applying the same analytical framework using real-time operational data could potentially reveal more pronounced spatial and temporal inequalities in accessibility than those suggested by timetable-based analyses.

Advances in large-scale open transport data and routing engines have made it possible to compute these travel times at increasingly fine spatial scales and across large geographical areas. In particular, the widespread adoption of the General Transit Feed Specification (GTFS) and the development of open-source routing engines such as OpenTripPlanner and R\textsuperscript{5} have enabled researchers to calculate multimodal origin–destination travel times efficiently and reproducibly using standardised data formats. These developments have substantially lowered the barriers to large-scale accessibility analysis and have helped establish reproducible computational workflows in transport data science research. In recent years, the increasing availability of real-time operational data offers new opportunities to capture the dynamics of transport systems and address the TTV and TTI problems. A growing body of work has begun to explore how real-time vehicle location data can be integrated with transit schedules to improve accessibility estimates \citep{wessel_accuracy_2019, braga_evaluating_2023, javanmard_using_2025, braga_understanding_2026}. However, implementing such approaches at large geographical scales remains computationally demanding with existing methodologies. Moreover, real-time transit datasets often contain incomplete or inconsistent information, requiring careful data cleaning and processing which often needs to be tailored for a specific dataset.

In this paper, we present \texttt{rt2gtfs}, an open-source Python package that implements a scalable framework for reconstructing empirically observed public transport timetables from real-time vehicle location data. Using national-scale real-time bus data from the UK Bus Open Data Service (BODS) \citep{department_for_transport_bus_2020}, the package links real-time vehicle positions to scheduled GTFS trips and stops, infers empirical stop-level arrival and departure times, and exports corrected GTFS timetable bundles. These observed timetables reflect how the network actually operated and can be used directly within existing routing engines to calculate observed travel times and accessibility. Building on the work of \citet{wessel_constructing_2017}, \texttt{rt2gtfs} prioritises computational efficiency and reproducibility so that corrected timetables can be produced repeatedly over long study periods and large geographical areas. Using six months of national-scale bus location data collected at high temporal frequency, we conduct a series of case studies to demonstrate how the package can be used to construct corrected timetables and quantify the differences between scheduled and observed accessibility. We further illustrate how the resulting dataset enables the analysis of travel time variability across space, revealing patterns that are not captured by conventional timetable-based approaches.

Our work is closely aligned with, and inspired by, the broader ecosystem of open-source software tools developed for urban research. In recent years, a growing number of tools have emerged to support different stages of the research lifecycle. These include tools that facilitate data acquisition from sources that are otherwise less accessible to researchers \citep{boeing_osmnx_2017, botta_packaging_2025, botta_rail_2025}, tools for data cleaning, processing, and format conversion \citep{morgan_uk2gtfs_2023, sato_city2graph_2026}, methods for computing or quantifying urban indicators \citep{biljecki_global_2022, mahajan_greenr_2024}, and analytical and modelling tools \citep{felix_reproducible_2025, botta_rail_2025, sevtsuk_madina_2025}. More recently, integrated platforms have also been developed that combine several of these functionalities within unified workflows \citep{ito_zensvi_2025, danish_citizen_2025}. Within the transport domain specifically, open-source tools have also been developed to enable large-scale accessibility modelling. These include language-specific wrappers for the R\textsuperscript{5} routing engine such as \textit{r5r} \citep{pereira_r5r_2021} and \textit{r5py} \citep{fink_r5py_2022}, as well as analytical libraries such as the \textit{accessibility} R package \citep{pereira_accessibility_2025}, which implements a range of accessibility measurement methods. These tools have played a crucial role in advancing reproducible research and have informed the design of \texttt{rt2gtfs}. In keeping with this open-science ethos, we provide the framework as a reusable Python package to enable reuse, replication, and further methodological development by other researchers. While \texttt{rt2gtfs} is currently optimised for UK BODS data, the underlying approach is based on GTFS and GTFS-Realtime standards and is in principle applicable to any GTFS and real-time vehicle location dataset that provide a linkable trip-level identifier.

The remainder of the paper is structured as follows. Section~\ref{sec:background} reviews existing research on travel time variability and the incorporation of real-time data into accessibility analysis. Section~\ref{sec:methodology} describes the proposed data processing and matching framework in detail. Section~\ref{sec:case} presents empirical case studies using England-wide bus data to illustrate the implications for accessibility measurement. Finally, Section~\ref{sec:conclusion} discusses limitations and directions for future research.

\section{Background}
\label{sec:background}
\subsection{The definition of travel time variability}
Travel Time Variability (TTV) describes a common phenomenon whereby travel time on recurrent journeys with the same origin, destination, and route may vary depending on several factors including time and date of departure and mode of travel. Numerous endeavours have attempted to dissect the concept of TTV while \citet{noland_travel_2002} proposed arguably the most widely adopted classification which includes three perspectives: vehicle-to-vehicle TTV, referring to variability in travel times between vehicles travelling on the same route at the same time; over the course of the day (within-day) TTV, referring to variability in travel times across different departure times within a single day; and day-to-day TTV, referring to variability observed for the same journey across different days. Research interest in public transport travel time variability (TTV) has been rising in recent years. The majority of existing work does not directly address TTV itself but instead mainly explores its impact on accessibility measurement with a strong emphasis on day-to-day TTV. A few exceptions quantified and examined within-day TTV directly and its spatial inequalities and socioeconomic implications \citep{farber_temporal_2014, wessel_accuracy_2019, chen_geography_2025}.

Before reviewing further TTV literature, it is worth mentioning that despite the term TTV being widely adopted in literature, several alternative terms have been used which is often due to the different causes of the variation in travel time. Some studies attempted to distinguish between them and TTV \citep{liu_realizable_2023, braga_evaluating_2023, javanmard_using_2025, braga_understanding_2026} but fundamentally they all describe some form of dynamic travel time phenomenon. We think it is important to discuss these different terms although we do not attempt to distinguish them, as often researchers use them synonymously or use one to mean another. Ultimately it is safe to assume that TTV would be the most generic and neutral term that covers all instances without revealing the direct cause of the phenomenon or excluding certain types of variability, despite it not always being used in this way.

\begin{itemize}

\item Travel Time Inaccuracy (TTI). TTI is not an alternative term to TTV but describes how observed travel times deviate from scheduled travel times \citep{braga_evaluating_2023}, which in turn results in systematic deviations of accessibility from a standard benchmark \citep{liu_realizable_2023}. Most public transport schedules are designed to make travel time between the same OD pairs relatively consistent across days in short periods of time, therefore TTI could result in TTV when the magnitude of this inaccuracy varies temporally. As a result, solving the issue of TTI is a predecessor to studying TTV. Existing literature mostly attributed TTI to traffic, weather, and operational conditions, which are the same causes associated with TTV \citep{liu_realizable_2023, braga_evaluating_2023}, while \citet{wessel_constructing_2017} added that in some cases TTI could also be due to agencies scheduling their services in a conservative manner to guarantee a higher adherence but in reality drivers may not strictly adhere to this schedule.

\item Travel Time Uncertainty (TTU). TTU is one of the most popular terms used in accessibility research to describe the dynamic nature of travel time. \citet{liu_realizable_2023} defined uncertainty in accessibility measurement as the stochastic variation in it, often due to the variation in travel time resulted from on-time performance and measurement error. One may assume TTU is one type of TTV which is not caused by expected fluctuation of travel demand, either within a day (i.e. peak vs off-peak) or seasonally. \citet{braga_evaluating_2023, braga_understanding_2026} conducted an extensive review of TTU literature but used the term day-to-day TTV in their own work, effectively treating TTU as synonymous with TTV. \citet{chen_reliable_2013, chen_measuring_2017, chen_understanding_2019, chen_evaluating_2020, chen_probabilistic_2026} maintained a constant interest in studying TTU over the past two decades by integrating it in the classical space-time prism (STP) model to measure accessibility more reliably, with the exception of \citet{chen_evaluating_2020} which uses floating catchment area (FCA) methods.

\item Travel Time Reliability (TTR). TTR is another popular term but also a very arbitrarily defined one, both by academia and government agencies \citep{taylor_travel_2013}, the latter of which often used it as a performance measurement for transport systems. Classical definition of TTR in academic literature describes it as the probability that travel between an origin-destination (OD) pair will perform adequately and be completed within an intended reasonable time \citet{nicholson_assessing_2003, bimpou_dynamic_2020}. The arbitrariness perhaps arises from this definition in that it is difficult to define or measure this `adequate performance'. \citet{australian_transport_council_national_2006} emphasised on-time performance for TTR assessment which is different from most other definitions that focus more on the degree of variation alone regardless of punctuality. For example, The UK Department for Transport defined TTR as the variation in journey times that
individuals are unable to predict \citep{department_for_transport_transport_2025}, which is effectively TTU as defined by \citet{liu_realizable_2023}. The Federal Highway Administration in the USA defined TTR as the consistency or dependability in travel times, as measured from day-to-day and/or across different times of day \citep{federal_highway_administration_travel_2006}, which is more closely aligned with TTV as it recognises the different types of variations. Parallel to defining TTR, the work of \citet{pu_analytic_2011} attempted to compare different TTR measurements, ranging from simple statistics (e.g. standard deviation and percentiles) to more sophisticated ones. This also provides valuable insight in measuring TTV for our work.

\end{itemize}

\subsection{Incorporating real-time data in travel time and accessibility calculation}
Travel time to destinations is central to most accessibility measurements as it provides a more meaningful representation of the cost borne by travellers than physical distance alone. Common indicators of accessibility include minimum travel time to the nearest essential service (e.g. the UK's Journey Time Statistics \citep{department_for_transport_journey_2019}); cumulative opportunity measures \citep{wachs_physical_1973}, which count the number of destinations reachable within a given travel time threshold; and a family of floating catchment area measures \citep{luo_measures_2003}, which incorporate both service supply and population demand to reflect spatial competition of limited resources.

Conventional accessibility studies rely on static, schedule-based data \citep{owen_modeling_2015}. Such representations assume that services operate exactly as timetabled. However, public transport systems are inherently subject to operational variability arising from traffic congestion, weather conditions, and various other factors causing service disruptions. As a result, schedule-based travel times may systematically underestimate observed and experienced travel times (TTI) and obscure temporal uncertainty. This limitation is particularly significant when accessibility metrics are used to assess spatial inequality, as variability and unreliability may disproportionately affect already disadvantaged populations. Existing literature has sought to address TTI by incorporating real-time data into travel time and accessibility calculations. Traditionally, these data will need to be manually collected by hand either by survey \citep{noland_simulating_1998} or actually travelling those trips and recording experienced travel times \citep{may_travel_1989}. This method continued on being implemented to this day \citep{duran-hormazabal_estimation_2016} but is costly and time-consuming to run at large geographical scale. Some studies have attempted to automate this using digital technologies such as the work of \citet{eliasson_forecasting_2009} using data from an automatic travel time measurement system which collects data through a camera system that captures images of number plates of individual vehicles as they enter and leave each road link. More recently, \citet{bimpou_dynamic_2020} used a novel data stream, the Google Distance Matrix API, to obtain travel times. However, the accuracy of such data is problematic as it is obtained by tracking Android devices through time and space and processed using an undisclosed algorithm.

The popularity of GPS technology made collecting real-time data much easier. Works such as \citet{mazloumi_using_2010, ma_modeling_2016, chepuri_examining_2018, rahman_analysis_2018} attempted to calculate transit travel time between route segments using GPS trajectory data and studied the distribution of these travel times and their temporal variation extensively. Some also attempted to integrate other forms of data such as smart card data to complement GPS data \citep{arbex_estimating_2020}. These studies tend to be applied only at route level on a few selected transit routes in a single city. Further research has also recognised the need to model door-to-door or origin-destination (OD) based travel times and attempted to apply some bespoke shortest path algorithm on transit networks \citep{farber_temporal_2014, langford_multi-modal_2016, chen_measuring_2017, chen_evaluating_2020, liu_realizable_2023, lee_social_2023}.

The increasing availability of large-scale public transport open data, particularly the arrival of General Transit Feed Specification (GTFS) standard for publishing transit timetables, alongside open-source routing engines such as OpenTripPlanner and R\textsuperscript{5} \citep{pereira_r5r_2021, fink_r5py_2022}, has made it easier to calculate multimodal OD travel times at scale. These developments have enabled accessibility analysis to be conducted with greater spatial granularity and computational efficiency than was previously possible. The more recent introduction of GTFS–Realtime (GTFS-RT) data was intended to address the limitations of schedule-based accessibility modelling by providing real-time updates on transit service operation. In principle, GTFS-RT supports stop-level arrival and departure times updates through \textit{TripUpdates} feeds. In practice, however, these updates frequently consist of predicted arrival and departure times \citet{webb_evaluating_2025} or delay offsets relative to the timetable \citep{auckland_transport_realtime_2026}, rather than ready-to-use empirically recorded stop events, and often require substantial manual processing before they can be used in travel time calculation and subsequent accessibility analysis. By contrast, the \textit{VehiclePositions} feed in GTFS-RT, which provides time-stamped geographic coordinates of vehicles (also known as Automated Vehicle Location (AVL) data), offers a more accurate representation of transit operations. The UK Bus Open Data Service (BODS) \textit{TripUpdates} specifically does not provide \textit{TripUpdates} in its GTFS-RT feed, and instead only supplies \textit{VehiclePositions} information. As a result, empirical stop-level arrival times cannot be directly obtained at national scale and must instead be reconstructed from high-frequency vehicle location data available. While this reconstruction approach allows for a more realistic representation of service performance, it introduces additional methodological complexity and computational demands. Nevertheless, incorporating such empirically derived travel times offers important opportunities to move beyond static representations of accessibility and to better capture the temporal variability that shapes the geography of transport-related inequalities.

Two main approaches for addressing TTI have since been developed which utilise this latest advancement in data standard and tools. \citet{wessel_constructing_2017} first developed a methodology to correct GTFS timetable data using AVL data which produces a retrospective timetable where the stop-level arrival times are modified to reflect their actual arrival times. The process involves matching vehicle positions with transit stops along the route it is operating on, and then use linear interpolation to derive the empirical arrival time at each bus stop. Applying this method on data from four cities in North America, \citet{wessel_accuracy_2019} found that TTI (i.e. using scheduled instead of observed empirical travel times) resulted in an inflation of accessibility by 5-15\% on average, and such overestimation shows strong spatial autocorrelation. The highlight of the Wessel methodology is that it seeks to maximise the accuracy of interpolated stop level arrival times using street network data, which also makes it computationally expensive to implement at large scale. What adds to the computational cost is that such process will have to be repeated for the entire study period to produce a revised timetable for each day, if day-to-day travel time variability is to be examined. To address this issue, \citet{braga_evaluating_2023} proposed a new method which constructs single corrected GTFS files across a study period, based on median and dispersion-based (85th percentile) travel times between transit stop segments. This results in two synthetic GTFS files where the travel time between each consecutive pair of stops are the median and 85th percentile travel times, calculated from all trips in the study period. They subsequently conducted two analyses using \textit{r5r} for calculating travel time and accessibility to jobs in Fortaleza, Brazil. Firstly, by comparing accessibility calculated from timetabled travel time and that from median observed travel times (i.e. TTI), they found that inaccuracy contributes to accessibility underestimation by 1.5\% on average, although in some area it could be over or underestimated by over 40\%. This is contrary to what \citet{wessel_accuracy_2019} found in their work which showed overwhelming overestimation. One explanation might be that, using observed timetable to calculate travel time makes the assumption that travellers have perfect knowledge of system operation when in reality they do not, as pointed out by \citet{liu_realizable_2023}. This results in a deflation of travel time which cannot be realised in real life scenarios. Secondly, by comparing accessibility calculated from median observed travel times and that of 85th percentile observed travel times, they found significant impact of TTV on accessibility estimation at 50\% on average, which is also unevenly distributed across the study space. Subsequent work used the same methods to explore accessibility to other destinations such as healthcare \citep{javanmard_using_2025} and found similar patterns. \citet{braga_understanding_2026} also sought to improve accessibility estimate by taking into account competition of resources. Such method greatly reduces computational cost but also has limitations in that it assumes all links in the public transport network operate under 50th or 85th percentile conditions, which may overestimate travel time. Our examination of the UK data found that this is not the case, and therefore our revised approach mainly took inspiration from the Wessel method and seeks to improve its computational efficiency while maintaining relative accuracy for our intended analysis. Such approach also allows us to directly examine the distribution and variability of travel times within a day and across days, which is currently understudied.

\section{Methodology}
\label{sec:methodology}

Here we briefly describe the method implemented in \texttt{rt2gtfs} to construct a corrected empirical timetable by matching real-time vehicle location data with scheduled timetable data. The approach builds on the method proposed by \citet{wessel_constructing_2017} and related work, but introduces several adaptations that substantially improve computational efficiency while preserving the accuracy required for large-scale accessibility analysis. We also highlight optimisations developed specifically for UK data which may or may not be applicable to other datasets.

The package is organised around a configurable processing workflow. Its main entry point, \texttt{construct\_observed\_gtfs()}, takes a target date and a \texttt{MatchingConfig} object specifying the locations of GTFS timetable and real-time dataset inputs, output directories, matching periods, search radius, time zone settings, interpolation options, whether additional timetable dates should be checked, and whether or not to output additional diagnostic files. This design separates the methodological parameters from the processing logic, making the workflow easier to reproduce, audit, and apply to multiple dates or regions. Detailed guidance and examples for configuring \texttt{MatchingConfig} and running the package are provided in the package documentation on GitHub. The principal stages of the package are described below: data collection, GTFS-RT parsing, vehicle-to-stop matching, stop time inference, interpolation of missing stop times, and export of a corrected GTFS bundle.

\subsection{Data collection}
We retrieved bus timetable and location data from the Bus Open Data Service (BODS) in GTFS (General Transit Feed Specification) and GTFS-Realtime (GTFS-RT) format respectively. It is worth noting that the data BODS supplies were originally stored in the UK's native data format, namely TransXChange for timetable and SIRI-VM for bus locations. The GTFS and GTFS-RT versions of the BODS data are said to be direct conversions from these native formats. We chose to use GTFS formats for our work for two reasons. Firstly, GTFS is the de-facto international standard for transit data first developed by Google. Using GTFS in \texttt{rt2gtfs} allows the package to be easily applicable in other parts of the world with minimal adaptations, promoting reproducibility. Secondly, the routing engine we utilise to calculate travel time, namely r\textsuperscript{5}py, requires GTFS format data as input. This is also the case for other popular open-source routing choices, such as OpenTripPlanner. Therefore, it is natural for us to collect GTFS data.

Like many public transport data services, BODS only provides access to current bus location data through live API feeds, with no archive of historical real-time vehicle location data available as of early 2026. Consequently, analysing travel time variability over extended periods requires the continuous collection and storage of these live data for later use. The \texttt{rt2gtfs} package includes data acquisition utilities for downloading vehicle location data from live feeds at user-defined intervals. By default, these utilities are configured for the UK BODS data, although users can specify alternative feed URLs where compatible data are available. For UK applications, the package also supports spatial filtering through a user-defined rectangular bounding box, allowing users to download a subset of the national BODS feed and reduce storage requirements when the full dataset is not needed. For the case study presented in this paper, we collect the full BODS bus location feed at least every 30 seconds, continuously throughout the year. In parallel, we also archive a complete copy of the UK’s GTFS bus timetable data on a daily basis. This mechanism enables the subsequent matching of observed vehicle locations to the corresponding published timetables, allowing construction of a corrected empirical bus timetable. Using this pipeline, we have been continuously collecting national-scale bus timetable and location data since February 2024, creating a growing archive of historical timetable and vehicle location records.


\subsection{Data preparation}

We begin our procedure by preparing the real-time data for the matching process. In its raw form, GTFS-RT is distributed as Protocolbuffer Binary Format (PBF) files, which are optimised for efficient data transmission but are not human readable. Following guidelines from the official GTFS-RT documentation \citep{general_transit_feed_specification_python_2025}, we therefore parse all downloaded GTFS-RT files into Python objects, and subsequently convert them into CSV formats. Note that a GTFS-RT feed can contain three types of messages: \textit{trip updates}, which reports real-time departure delays of trips; \textit{vehicle positions}, which reports real-time position of vehicles; \textit{alerts}, which reports real-time alerts. The UK BODS data provides only vehicle position feeds, which has important implications for the design of our matching procedure, as stop-level delay information is not directly available, and therefore needs to be inferred by matching vehicle location data with stops data in timetable data. This also implies that our matching procedure can be easily adapted to other forms of real-time vehicle location data, not only GTFS-RT, provided that the data include a trip-level identifier key that allows observations to be linked to a corresponding timetable (further details below).

All vehicle position records collected within the same day are merged into a single CSV dataset, after which duplicate records are removed. On a typical day, approximately 50\% of the raw observations are duplicates (same vehicle at the same position and timestamp), reflecting the fact that vehicles are frequently stationary between successive API calls, for example while dwelling at stops or waiting at depots. Removing these duplicates substantially reduces data volume while preserving the full spatio-temporal trajectory of each vehicle, preparing the dataset for efficient and reliable matching in the next stage of the package workflow.

\subsection{Matching vehicle locations to stops}
With data preparation complete, we proceed to matching real-time vehicle locations to the corresponding GTFS timetable. For this matching to be possible, each vehicle position record must contain a valid \textit{trip\_id} input, which serves as the key linking real-time observations to the corresponding timetabled \textit{trip} \footnote{In a GTFS feed, a trip represents a single, scheduled run of a vehicle along a specific route and direction. Each trip corresponds to one instance of a service along a \textit{route} operating at a particular time on a given day, defined by a unique \textit{trip\_id}, and linked to a \textit{route\_id}, a \textit{service\_id} (specifying the days on which it runs), and a \textit{shape\_id} (describing the spatial path it follows). A trip is realised operationally when a physical vehicle is assigned to it and travels along its sequence of stops according to the timetable.} within the GTFS bundle. This identifier allows each observed vehicle trajectory to be associated with a specific scheduled \textit{trip}, alongside the \textit{route} it is operating, \textit{stop sequence}, and stop-level scheduled \textit{arrival and departure time}, forming the basis for correcting the timetable with real-time information. Like many large scale datasets, BODS data contain substantial missing information. In particular, not all vehicle position records include a valid \textit{trip\_id}, which in principle prevents them from being directly matched to the scheduled GTFS timetable. Some previous studies have attempted to mitigate this issue by using the \textit{vehicle\_id} field to infer or interpolate missing \textit{trip\_id} values \citep{open_innovations_transport_2024, open_innovations_tracking_2025}. In our experiments, however, these approaches yielded only marginal improvements in data completeness.

\begin{figure}[H]
    \centering
    \includegraphics[scale=0.45]{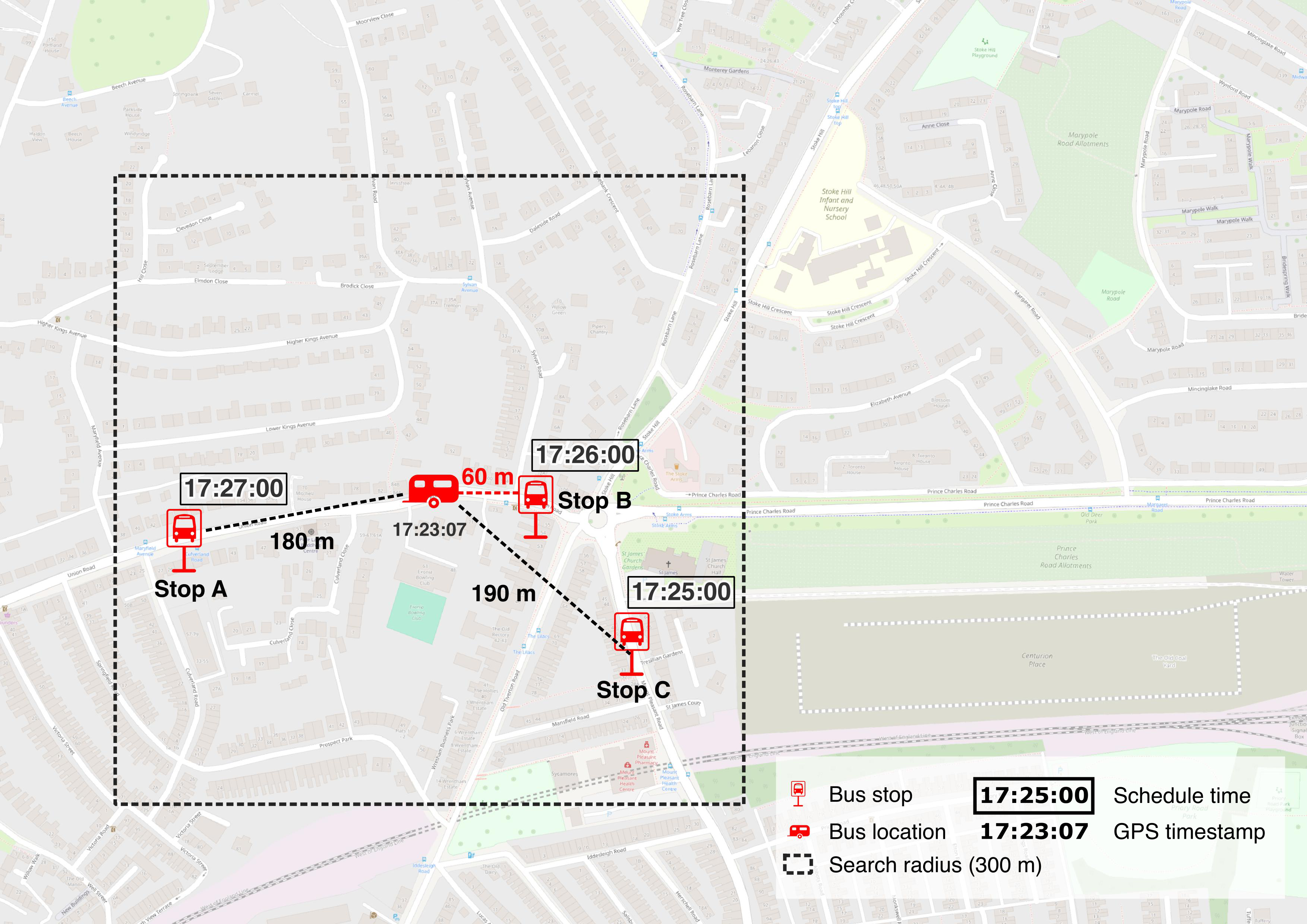}
    \caption{Map showing the matching procedure. Candidate Stops A, B, and C were identified during the initial coarse search within a rough 300-metre radius of the recorded vehicle position at 17:23:07. The exact distance between the vehicle position and the three candidate stops were then calculated, with Stop B being identified as the closest match at the shortest distance of 60 metres. Basemap tiles © OpenStreetMap contributors.}
    \label{fig:matching}
\end{figure}

Next, we process each matchable vehicle position record. For each observation, we identify the corresponding scheduled \textit{trip} in the GTFS feed and retrieve its associated \textit{route} and ordered list of \textit{stops}, each with known geographic coordinates (longitude and latitude). To efficiently locate the stop most likely associated with a given vehicle position, we first performed a coarse spatial search along the route to identify candidate stops within a bounding box centred on the vehicle's reported location (Figure~\ref{fig:matching}). For a vehicle position \(p\), a stop \(s\) on route \(r\) was retained as a candidate if
\[
\left| \phi_s - \phi_p \right| < \delta
\quad \text{and} \quad
\left| (\lambda_s - \lambda_p)\cos(\phi_p) \right| < \delta,
\]
where \(\phi\) and \(\lambda\) denote latitude and longitude respectively, and \(\delta\) is a small angular threshold corresponding which can be customised (300 metres used in our work). The cosine adjustment accounts for the convergence of meridians with latitude. This coarse filter substantially reduces computational cost by avoiding the calculation of exact distances to all stops along the route. When one or more candidate stops are identified, we then compute the precise distance between the vehicle and each candidate stop and select the closest stop as the most likely match. For each such match, we record the matched \textit{stop\_id}, the corresponding distance, and the timestamp of the vehicle position observation. Repeating this procedure across all records produces a large dataset of vehicle stop matchings over time.

\subsection{Inferring empirical stop times}
\label{sec:infer}
These matchings are then used to infer observed stop times. Because a vehicle may be matched to the same stop multiple times as it approaches and departs, we retain only the match with the minimum distance for each stop–vehicle pair. The timestamp of this closest observation is taken as the observed empirical arrival time and is used to replace the scheduled time in the GTFS \textit{stop\_times.txt} file. GTFS \textit{stop\_times.txt} contains separate \textit{arrival\_time} and \textit{departure\_time} fields. However, in the BODS GTFS feeds used in this study, these two fields are typically identical for bus stops, meaning that scheduled dwell time is not explicitly represented. \texttt{rt2gtfs} therefore does not infer separate stop-level observed arrival and departure times. Instead, the same inferred observed stop time is written to both fields. This preserves consistency with the structure of the input timetable data while producing a corrected GTFS feed that remains compatible with standard routing engines. It is worth noting that our approach does not estimate the exact moment when a vehicle comes to rest at a stop, but rather the moment when it is closest to that stop. This represents a deliberate trade-off compared with the method of \citet{wessel_constructing_2017}, which uses network-based interpolation to obtain more precise arrival and departure times at the cost of substantially greater computational complexity. However, when real-time vehicle positions are collected at high temporal frequency, the difference between the closest recorded observation and the true stop event is likely to be relatively small in many cases. For the purposes of accessibility analysis, where the primary concern is the cumulative impact of operational delays on end-to-end travel times rather than punctuality at individual stops, this approximation provides a scalable and sufficiently accurate solution.

\subsection{Interpolating missing stop times}

Unfortunately, not all stops along a bus route can be matched with live vehicle position observations. This can occur when vehicles pass through stops without dwelling, when vehicles travel quickly between stops, or when the sampling interval of the real-time feed misses the moment of closest approach. In rare cases, large segments of GPS data may be lost when vehicles travel through rural areas with weak mobile signal coverage, preventing onboard equipment from transmitting data to the server. 

Before interpolating missing stop times, \texttt{rt2gtfs} applies a temporal consistency check to the matched observations within each trip. Because stop times should follow the order of the scheduled stop sequence, cases where an earlier stop is assigned a later observed time than a subsequent stop are treated as inconsistent and removed from the set of interpolation anchors. This step reduces the risk that erroneous spatial matches, noisy GPS observations, or irregular vehicle position updates propagate into the reconstructed timetable.

Missing stop times are then interpolated using the nearest successfully matched stops along the same trip. When both upstream and downstream matched stops are available, we linearly interpolate the missing stop time using the scheduled timetable as a reference. When missing stops occur at the beginning or end of a trip, such that only one neighbouring matched stop is available, we use one-sided extrapolation. We recognise that this extrapolation approach is less robust than interpolation, as it assumes that the deviation between scheduled and observed time at the nearest matched stop remains constant across neighbouring stops. However, this approach preserves the temporal structure of the scheduled timetable and enables complete stop-level arrival/departure times to be reconstructed.

Formally, let $n$ denote a stop with scheduled arrival time $t_n$ and estimated observed arrival time $\hat{r}_n$. Let $p$ and $q$ denote the nearest upstream and downstream stops that have been successfully matched with real-time observations. Their scheduled arrival times are $t_p$ and $t_q$, while their observed arrival times are $r_p$ and $r_q$, respectively.

\begin{equation}
\hat{r}_n =
r_p +
\frac{t_n - t_p}{t_q - t_p}
\left(r_q - r_p\right),
\qquad \text{if both } r_p \text{ and } r_q \text{ are available.}
\label{eq:interpolation}
\end{equation}

\begin{equation}
\hat{r}_n =
t_n + \left(r_q - t_q\right),
\qquad \text{if } r_p \text{ is unavailable and } r_q \text{ is available.}
\label{eq:backward_extrapolation}
\end{equation}

\begin{equation}
\hat{r}_n =
t_n + \left(r_p - t_p\right),
\qquad \text{if } r_q \text{ is unavailable and } r_p \text{ is available.}
\label{eq:forward_extrapolation}
\end{equation}

Equation~\ref{eq:interpolation} linearly maps the scheduled temporal position of an unmatched stop within the interval $[t_p, t_q]$ to the corresponding interval in observed real time $[r_p, r_q]$. Table \ref{tab:interpolation_example} shows an illustration of this interpolation process. Equations~\ref{eq:backward_extrapolation} and \ref{eq:forward_extrapolation} are used for stops at the beginning and end of a trip, respectively, where only one neighbouring matched stop is available. In these cases, the scheduled stop time is shifted by the observed deviation between timetable and real time at the nearest matched stop. For example, stops on a trip may only be matched starting from the second stop on the route. If the vehicle arrives at Stop 2 with a two-minute delay, our extrapolation estimates the observed arrival time at Stop 1 by assuming that the vehicle also left that stop two minutes late (Table \ref{tab:extrapolation_example_start}). Similarly, if the vehicle arrives at Stop 29 (second-to-last) with a two-minute delay, our extrapolation estimates the observed arrival time at Stop 30 by assuming that the vehicle also arrived at that stop two minutes late (Table \ref{tab:extrapolation_example_end}). This assumption is unlikely to always hold in reality, as delays can accumulate or dissipate between stops depending on traffic conditions, passenger boarding times, or operational adjustments by the driver. Nevertheless, in the absence of real-time observations at those stops, this approach provides a reasonable approximation that preserves the temporal structure of the resulting timetable.

\begin{table}[H]
\centering
\begin{tabular}{cccc}
\toprule
Stop Sequence & Scheduled & Matched & Interpolated \\
\midrule
1 & 14:25:00 & 14:27:02 & 14:27:02 \\
2 & 14:26:00 & NaN & 14:28:34 \\
3 & 14:27:00 & 14:30:06 & 14:30:06 \\
\bottomrule
\end{tabular}
\caption{Illustrative example of interpolation used to estimate missing stop arrival times. In this example, Stop 2 could not be matched so we used Equation \ref{eq:interpolation} for the interpolated time.}
\label{tab:interpolation_example}
\end{table}

\begin{table}[H]
\centering
\begin{tabular}{cccc}
\toprule
Stop Sequence & Scheduled & Matched & Interpolated \\
\midrule
1 & 14:25:00 & NaN & 14:27:00 \\
2 & 14:26:00 & 14:28:00 & 14:28:00 \\
\bottomrule
\end{tabular}
\caption{Illustrative example of extrapolation (Equation \ref{eq:backward_extrapolation}) used to estimate missing stop arrival times at the beginning of a trip.}
\label{tab:extrapolation_example_start}
\end{table}

\begin{table}[H]
\centering
\begin{tabular}{cccc}
\toprule
Stop Sequence & Scheduled & Matched & Interpolated \\
\midrule
29 & 15:23:00 & 15:25:00 & 15:25:00 \\
30 & 15:24:00 & NaN & 15:26:00 \\
\bottomrule
\end{tabular}
\caption{Illustrative example of extrapolation (Equation \ref{eq:forward_extrapolation}) used to estimate missing stop arrival times at the end of a trip.}
\label{tab:extrapolation_example_end}
\end{table}

\subsection{Finalisation and optimisation}
At the end of this procedure, \texttt{rt2gtfs} writes a corrected GTFS bundle containing the services that were matched as operated and their reconstructed stop-level arrival/departure times. The output preserves the GTFS structure expected by routing engines, allowing the observed timetable to be used directly for calculating observed travel times.

Optimisations were implemented in \texttt{rt2gtfs}, with some designed specifically for UK BODS data. First, because a substantial number of vehicle position records lack valid \textit{trip\_ids}, we sought to maximise the usefulness of those records that do contain this identifier. In practice, we found that some \textit{trip\_ids} present in the operational data could not be located in the corresponding GTFS timetable for the same day. This is likely due to the bus operators not updating their GTFS timetable data or the \textit{trip\_id} used in their operation. To address this, we extended the matching process beyond the timetable for the observation date to include GTFS feeds from a specified number of days before and after the target date. As this increases the computational time required for matching, the number of additional days included was made configurable in the package. In our experiments, extending the matching window to seven days before and after the observation date increased the proportion of matchable vehicle position records by up to 20\% on some days, representing a substantial improvement.

\subsection{Outputting matching details and diagnostics}

In addition to producing empirically corrected GTFS timetables, \texttt{rt2gtfs} can generate stop-level matching details and diagnostic outputs that summarise the matching process. These outputs are designed to help users evaluate the quality and completeness of the reconstructed timetable before using it in routing or accessibility analysis.

\begin{figure}[H]
    \centering
    \includegraphics[width=1\linewidth]{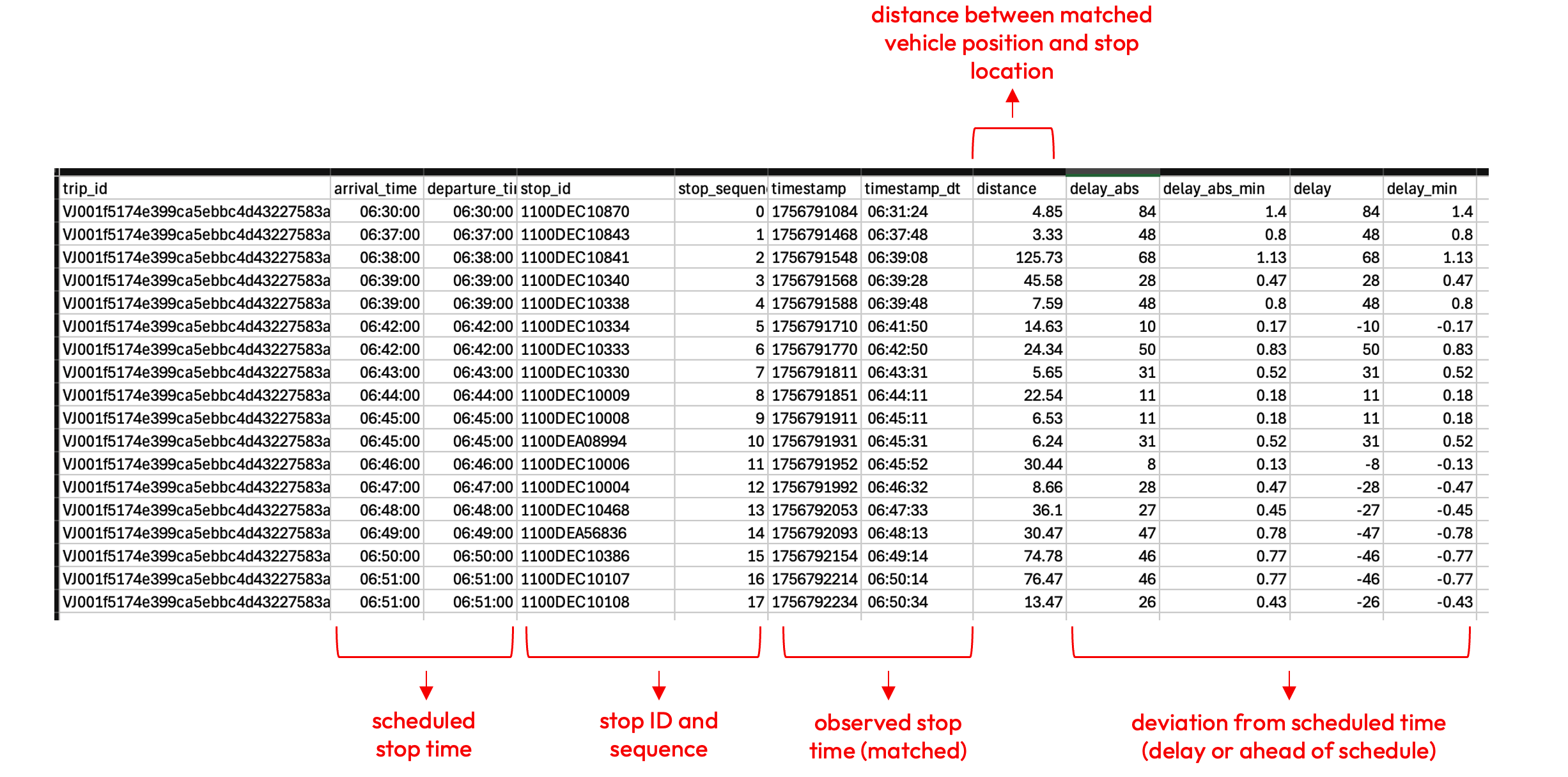}
    \caption{Example of the optional matching detail output generated by \texttt{rt2gtfs}. Each row represents a scheduled stop on a specific bus journey along a specific bus route. The table reports, from left to right, the scheduled arrival/departure time, stop ID, stop sequence, matched empirical stop arrival/departure time, distance between the stop and the nearest matched vehicle location, and the time difference between the scheduled stop time and the matched timestamp.}
    \label{fig:matching_details}
\end{figure}

For stop-level matching details, when enabled through the package configuration, \texttt{rt2gtfs} outputs a table detailing the relationship between scheduled stops and matched vehicle positions (Figure \ref{fig:matching_details}). This table includes the straight-line distance between each bus stop and the closest matched vehicle location, as well as the time difference between the scheduled stop time and the matched vehicle timestamp. The time difference is reported both as a signed value, where positive values indicate delay and negative values indicate running ahead of schedule, and as an absolute value. It is important that users do not assume that this is a precise delay measurement, and therefore should exercise caution when using these outputs directly for punctuality analysis. The matched timestamp represents the time at which a vehicle location observation was closest to a nearby scheduled stop, rather than an empirically recorded arrival or departure event at that stop, as outlined in Section~\ref{sec:infer}. Yet, these outputs can provide useful insights on service performance and matching plausibility, but they should not be treated as a direct substitute for validated stop-level arrival or departure records.

\begin{figure}[H]
    \centering
    \includegraphics[width=1\linewidth]{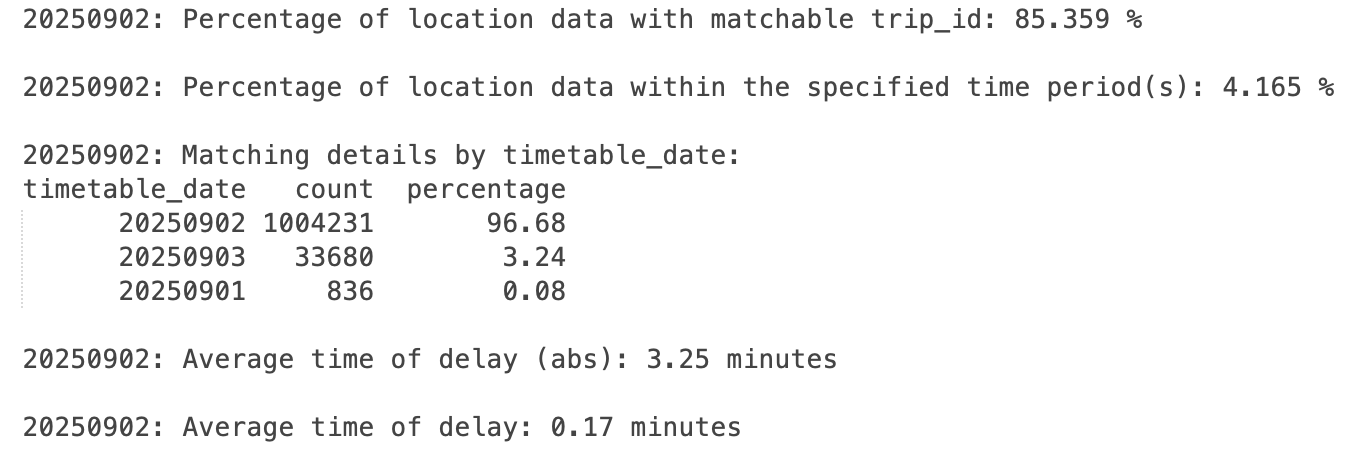}
    \caption{Example of the optional matching diagnostics output generated by \texttt{rt2gtfs}. The output summarises the performance of the matching process, including the proportion of real-time vehicle location records that could be matched to scheduled GTFS records, the timetable dates used for matching when additional archived GTFS feeds from other dates are considered and their corresponding proportions, and aggregate delay-related statistics derived from differences between matched vehicle timestamps and scheduled stop times.}
    \label{fig:matching_diag}
\end{figure}

\texttt{rt2gtfs} can also generate summary statistics for the matching process. Not all real-time vehicle location observations can be matched to scheduled GTFS records. This may occur, for example, when the observed \textit{trip\_id} is absent from the corresponding timetable feed, or when the vehicle position falls outside the spatial search radius of candidate stops on the scheduled trip. In \texttt{rt2gtfs}, this search radius is user-configurable and defaults to 300 metres. The summary statistics include the proportion of real-time vehicle location observations that can be matched to scheduled GTFS records, as well as the timetable dates to which real-time observations are matched and their relative proportions when matching across multiple days is enabled. The package can also report aggregate delay-related statistics by comparing matched vehicle timestamps with the corresponding scheduled stop times. Figure~\ref{fig:matching_diag} shows an example of the diagnostic output.

Beyond supporting the construction of corrected GTFS timetables, these outputs can enable a range of additional analyses of public transport operations. For example, the proportion of real-time observations successfully matched to scheduled records can be used to assess data completeness and feed quality across dates, regions, or operators. The distribution of matched timetable dates can help identify inconsistencies between real-time feeds and published timetables. Delay-related summaries can also provide indicative evidence of service reliability, although, as noted above, they should be interpreted with caution when used for punctuality analysis. These outputs therefore provide both an analytical resource and an important quality control step. Reporting matching statistics allows users to identify days, regions, or feeds with poor match quality, assess whether additional timetable dates should be considered, and evaluate the reliability of the corrected timetable, etc.

\section{Software and data availability}
\label{sec:software}

The methodology described in this paper is implemented in the open-source Python package \texttt{rt2gtfs}. The package is available from GitHub at \href{https://github.com/kevinwinsper/rt2gtfs}{https://github.com/kevinwinsper/rt2gtfs}. The repository includes documentation, a tutorial, and example workflows demonstrating how to configure and run the package. \texttt{rt2gtfs} provides reusable functions for downloading GTFS and GTFS-RT data, parsing vehicle position feeds, matching observations to scheduled trips and stops, interpolating missing stop-level times, and exporting corrected GTFS bundles.

The case studies presented here use data from the UK Bus Open Data Service (BODS). BODS provides live access to current vehicle location feeds, but not a historical archive of past vehicle positions. The historical dataset used in this study was therefore constructed through continuous collection by the authors. Sample data for running and testing the package are available through Zenodo at \href{https://doi.org/10.5281/zenodo.19889312}{https://doi.org/10.5281/zenodo.19889312}. Additional processed data can be made available where licensing, size, and storage constraints allow.

\section{Case Studies}
\label{sec:case}
To demonstrate how the corrected timetable reveals differences between scheduled and observed accessibility, as well as day-to-day variability over extended periods, we present a set of case studies using England-wide BODS data from May to October 2025.

The \texttt{rt2gtfs} package was applied to each day in the study period, producing a corrected timetable that was passed to r\textsuperscript{5}py, a Python interface to the R\textsuperscript{5} routing engine for fast multimodal routing \citep{fink_r5py_2022}. Bus travel times from each Lower layer Super Output Area (LSOA) \footnote{Lower Layer Super Output Areas (LSOAs) are small statistical geographic units used in England and Wales for reporting census and administrative data, each containing roughly 1,500 residents.} to selected essential services were computed for multiple departure times, and summarised using their mean and standard deviation to characterise typical travel time and travel time variability (TTV).

\subsection{Case study 1: Within-day travel time and TTV for hospital access}
\begin{figure}[H]
    \centering
    \includegraphics[scale=0.45]{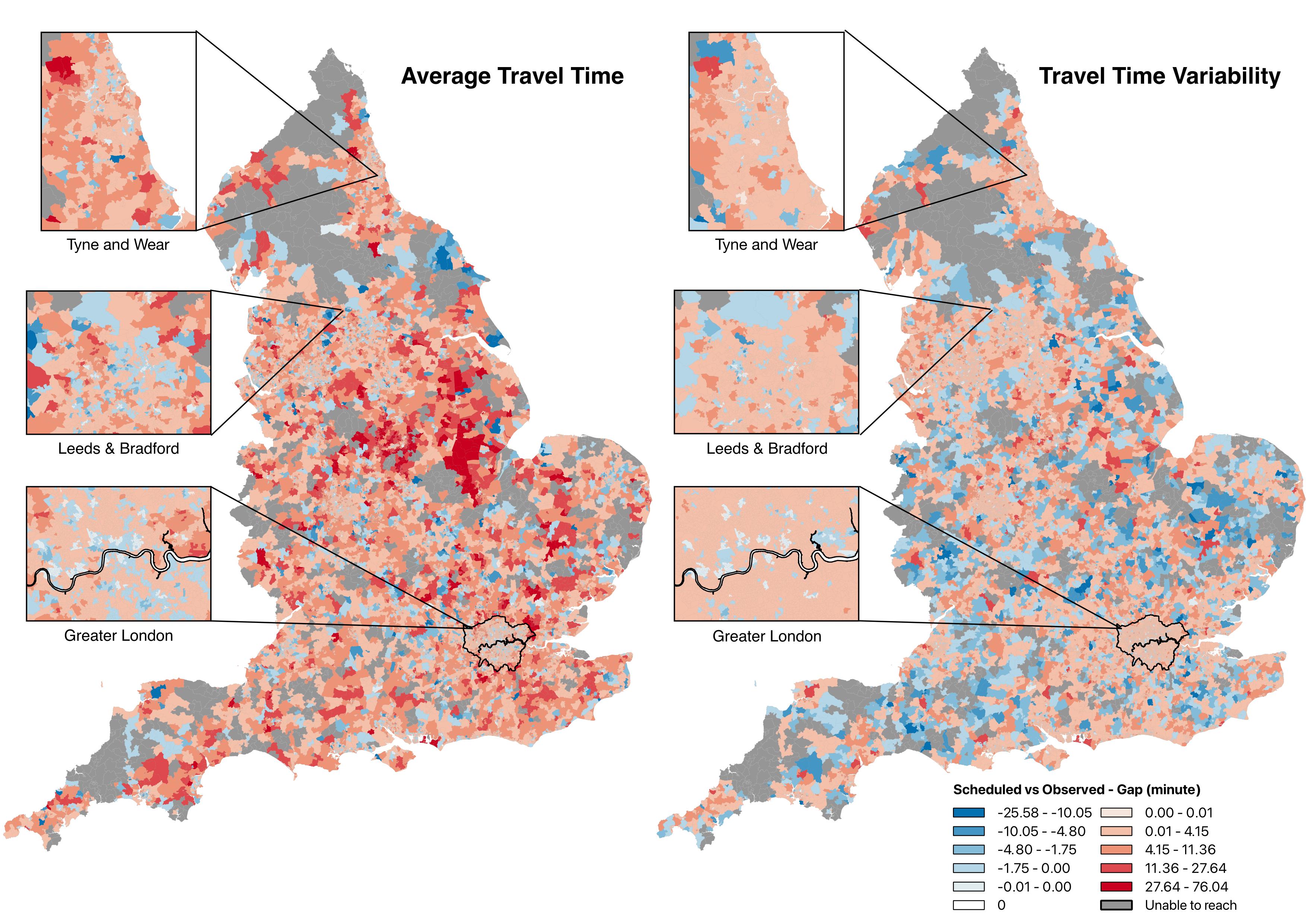}
    \caption{Maps of the \textbf{differences} between scheduled and observed average travel time and travel time variability (TTV) for journeys from each LSOA in England to the nearest \textbf{hospital}. \textbf{Red} indicates observed travel time and TTV are higher than scheduled while \textbf{blue} indicates the opposite. LSOAs shown in \textbf{grey} are unable to reach a hospital by bus within two hours.}
    \label{fig:hosp_diff}
\end{figure}

We first analysed within-day travel time and TTV for journeys to hospitals \footnote{The hospitals used were from the \href{https://digital.nhs.uk/data-and-information/publications/statistical/estates-returns-information-collection/summary-page-and-dataset-for-eric-2024-25}{Estates Return Information Collection (ERIC) Site} Data for year 2024/25 published by NHS Digital \citep{nhs_england_digital_estates_2024}. We used the same selection criteria as the DfT Journey Time Statistics to include only the NHS `general' hospitals. For details: \href{https://www.gov.uk/government/publications/journey-time-statistics-guidance/journey-time-statistics-notes-and-definitions-2019}{https://www.gov.uk/government/publications/journey-time-statistics-guidance/journey-time-statistics-notes-and-definitions-2019}}. Hourly travel times from 7:00 to 17:00 were calculated from each LSOA to all hospitals, retaining the shortest time in each hour. Scheduled travel times and TTV for a representative weekday (the second Tuesday of each month) were used as a benchmark, while observed travel times and TTV from July to October 2025 were averaged across weekdays. Figure \ref{fig:hosp_diff} maps the resulting differences between scheduled and observed average travel time and TTV. Across the study period, 79.3\% of LSOAs experienced longer observed average travel times than scheduled (79.9\% for urban and 76.2\% for rural), while 81.6\% exhibited higher observed TTV (86.3\% for urban and 57.7\% for rural). Note that here we consider a travel time to be longer if the difference between observed and scheduled travel time is strictly positive, so even a small difference of 1 second contributes to this. We report this statistic only for demonstrative purposes of the type of analysis our work may enable researchers to do; in policy and operational settings, it is more likely to be appropriate to only consider differences which are larger (or smaller) than a given threshold (e.g. a bus may only be considered to be late if the difference is larger than 60 seconds).

\subsection{Case study 2: Day-to-day travel time and TTV for town centre access}
\begin{figure}[H]
    \centering
    \includegraphics[scale=0.16]{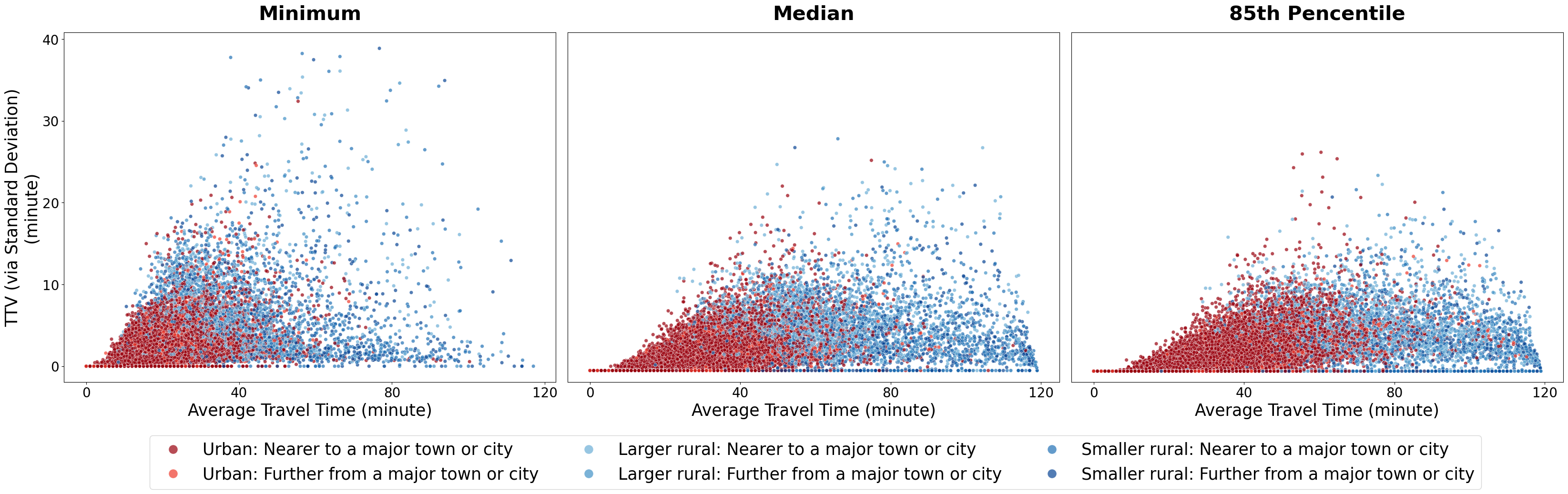}
    \caption{\textbf{Scatterplot} of travel time variability (TTV) versus average travel time for trips from each LSOA in England to the nearby town centre, with points color-coded by settlement type (urban/rural). Minimum, median, and 85th percentile travel times were calculated for travel times between 8:00 and 10:00 on each weekday from May to October 2025. }
    \label{fig:town_centre}
\end{figure}

We next examined day-to-day travel time variability during the morning peak. Travel times from each LSOA to nearby town centres \footnote{We used locations of English town centres in 2004 from the DLUHC (Department for Levelling Up, Housing and Communities) Town Centres and retail planning statistics for England and Wales.} were calculated at one-minute intervals between 08:00 and 10:00 on each weekday across the study period, producing 120 travel time estimates per origin–destination pair per day. Minimum, median, and 85th percentile travel times were extracted for each two-hour window. Day-to-day TTV is defined as the standard deviation of these travel times, while average travel time is their mean. Figure \ref{fig:town_centre} reveals a general positive correlation between TTV and average travel time, with urban areas exhibiting lower values for both metrics, while rural areas tend to show higher variability and longer travel times. Overall, the distribution highlights the urban-rural divide in accessibility.

\subsection{Case study 3: GP accessibility when accounting for TTV and competition for resources}

Last but not least, we examine access to general practitioner (GP) services in England using two commonly used accessibility measures: cumulative opportunities and the two-step floating catchment area (2SFCA) method \citep{luo_measures_2003}. While cumulative opportunity measures count the number of services reachable within a given travel-time threshold, 2SFCA further accounts for the balance between healthcare supply and population demand within catchment areas.

Observed public transport travel times were calculated from every LSOA in England to all GP practices, using corrected timetable outputs from the \texttt{rt2gtfs} package. Travel times were calculated separately for each hourly departure time from 07:00 to 17:00 on every weekday between May and October 2025. Each departure hour on each weekday was therefore treated as a separate travel time scenario. For each scenario, the resulting LSOA to GP travel time matrix was used as the cost input to calculate cumulative opportunity and 2SFCA accessibility scores using Python package \texttt{access} \citep{saxon_open_2021}. For each LSOA, we calculate average accessibility as the mean accessibility score across all departure hours and weekdays. We also calculate an accessibility variability measure in two stages. We first calculated the standard deviation of accessibility scores across departure hours within each weekday, resulting in a within-day accessibility variability measure for each weekday. We then averaged these daily standard deviations across all weekdays to obtain an overall measure of typical accessibility variability.

\begin{figure}[H]
    \centering
    \includegraphics[width=1\linewidth]{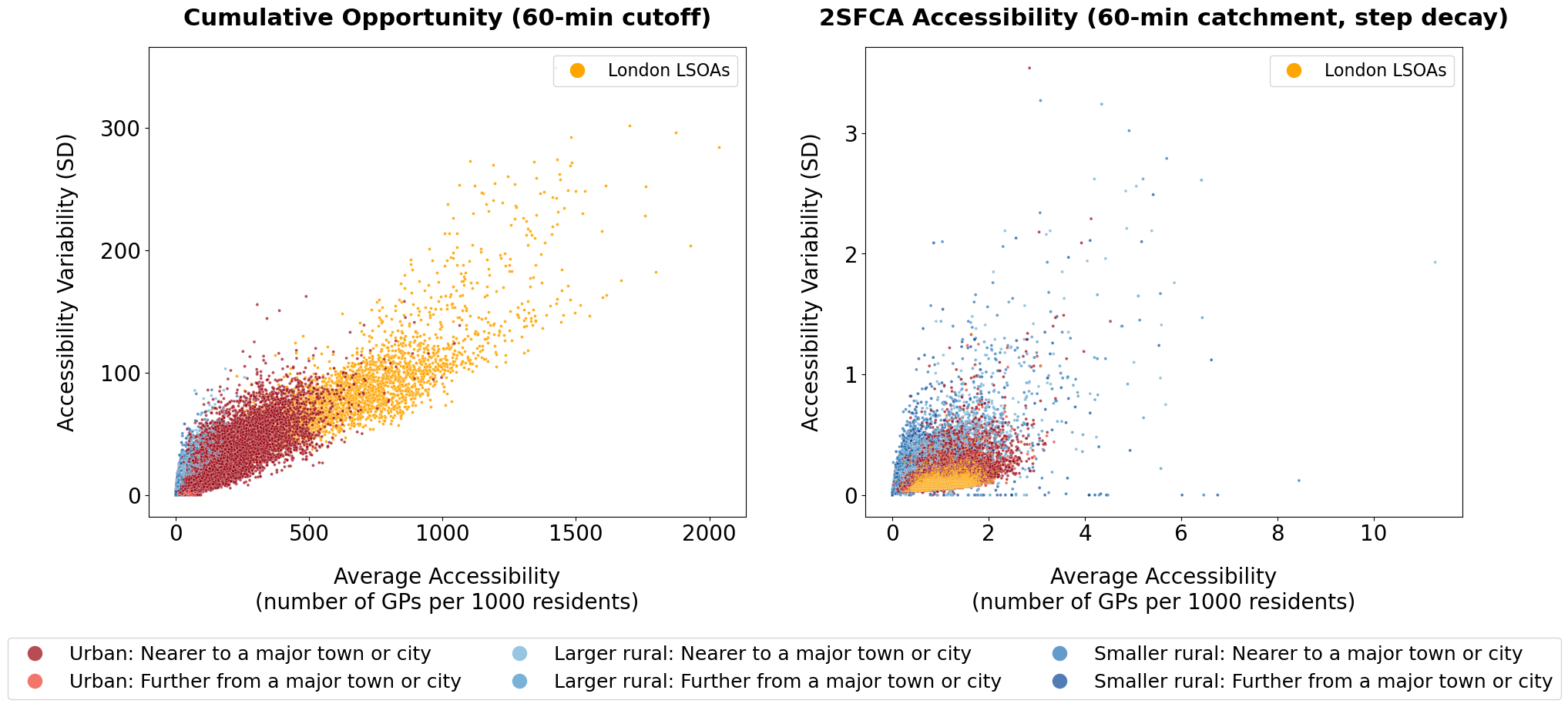}
    \caption{Comparison of average GP accessibility and accessibility variability across LSOAs in England using two accessibility measures: cumulative opportunity with a 60-minute cutoff and 2SFCA accessibility with a 60-minute catchment and step decay. Each point represents an LSOA, coloured by settlement type, with London LSOAs highlighted in yellow. The cumulative opportunity measure produces substantially higher accessibility values in London, reflecting the high density of GP practices, whereas the 2SFCA measure accounts for supply-demand competition and produces a more realistic distribution. In both measures, accessibility variability is represented by the standard deviation across observed travel times across a six-month period.}
    \label{fig:gp_acc}
\end{figure}

The results reveal differences in average accessibility and accessibility variability across settlement types when competition for resources is taken into account. Figure \ref{fig:gp_acc} (right) reveals that urban and rural areas show broadly similar average levels of GP accessibility when supply and demand are accounted for, but rural areas experience higher variability overall. This suggests that rural populations may be more exposed to and affected by fluctuations in public transport travel times, even when their average accessibility appears comparable to that of urban areas. This case study therefore illustrates how \texttt{rt2gtfs} can support more nuanced assessments of healthcare accessibility by linking observed public transport performance with spatial measures of service provision and population demand.

Our results (Figure \ref{fig:gp_acc} - left) also demonstrate that cumulative opportunity measures tend to produce high accessibility scores in dense urban areas, particularly London, where the concentration of GP practices is high. However, these values may overstate effective access because they do not account for competition from surrounding populations. In contrast, the 2SFCA method provides a more realistic estimate of GP accessibility by considering both the supply of GP services and the population competing for them.

\section{Conclusion}
\label{sec:conclusion}
This paper has presented \texttt{rt2gtfs}, an open-source Python package for constructing empirical measures of bus transport travel time and accessibility and their dynamics using national-scale real-time operational data. By matching real-time vehicle location feeds with scheduled GTFS timetables, the package implements a computationally efficient procedure that reconstructs observed stop-level arrival/departure times and produces a corrected, routable timetable reflecting how the bus network actually operated. Although the approach involves certain simplifications for efficiency gains, it produces a reconstructed timetable that is well suited for large-scale accessibility analysis. Applying this framework to half a year of BODS data for England, we computed travel times from every LSOA to selected essential services and derived measures of travel time variability that reveal pronounced spatial and temporal inequalities of accessibility and differences between scheduled and observed accessibility.

While this approach provides a substantial advance over timetable-based accessibility analysis, it is subject to several important limitations. Firstly, the analysis focuses exclusively on bus services and does not include other public transport modes. Although this does not capture the full public transport network, buses represent the most widely available form of public transport across much of the country, particularly in rural and smaller urban areas where alternative modes are limited. Buses are also generally the most affordable public transport option, and therefore play a central role in shaping everyday accessibility to a wide range of essential services.
Secondly, the observed travel times reconstructed in this study are retrospective and represent what would only be achievable if travellers had perfect knowledge of service delays and disruptions at the time of travel. In reality, travellers make decisions under uncertainty and incomplete information (often only paper-based static timetables), meaning that the accessibility calculated using the corrected timetables produced by \texttt{rt2gtfs} may not necessarily be realisable in practice. However, as real-time passenger information systems become increasingly widespread through mobile applications, on-board announcements, and digital displays at bus stops, travellers are increasingly gaining more awareness of accurate and timely travel advice. This trend suggests that empirically observed travel times like the ones we calculated are becoming a closer approximation to what informed travellers can actually experience, particularly in urban areas and on high-frequency routes. From a planning and policy perspective, our measures therefore provide an upper bound on achievable accessibility under real-time information, offering a benchmark against which the performance and reliability of public transport systems can be assessed. 
Thirdly, not all vehicle position records can be matched to scheduled trips because of missing \textit{trip\_id} information in the BODS feed. As a result, some areas may appear to experience poorer accessibility not because services are genuinely less frequent or reliable, but because parts of their operations are poorly recorded. While we have no direct control over the quality of BODS data, future work will seek to further improve matching rates through more sophisticated \textit{trip\_id} interpolation and inference methods, enabling even more complete and realistic representations of public transport accessibility. Fourth, the inference of missing stop-level times introduces uncertainty. The interpolation and extrapolation procedures assume that delays change smoothly between matched stops, or remain constant for unmatched stops at the beginning and end of a trip. In practice, delays may accumulate or dissipate unevenly due to traffic, passenger boarding activity, or driver behaviour. The inferred times should therefore be understood as approximations that preserve the temporal structure of the timetable, rather than exact observations at every stop.

Overall, \texttt{rt2gtfs} provides a computationally efficient and scalable software framework for national-level analyses, offering a practical approach for incorporating real-time data into more reliable and realistic accessibility assessments and providing an improved evidence base for service evaluation and transport planning.

\section*{Acknowledgement}
This work was supported by UK Research and Innovation (EP/S022074/1 UKRI Centre for Doctoral Training in Environmental Intelligence: Data Science \& AI for Sustainable Futures).

For the purpose of open access, the author has applied a Creative Commons Attribution (CC BY) licence to any Author Accepted Manuscript version arising from this submission.

The authors thank Prof Stewart Barr (University of Exeter) for their valuable input and insights, which improved this work.

\section*{Declaration of generative AI and AI-assisted technologies in the writing process}

During the preparation of this work the authors used ChatGPT by OpenAI and Claude by Anthropic to provide feedback on early drafts and to support language editing for clarity and readability. After using this tool, the authors reviewed and edited the content as needed and take full responsibility for the content of the published article.

\bibliographystyle{elsarticle-harv}
\bibliography{references}

\end{document}